\documentclass[aps,prl,reprint,amssymb,amsmath,superscriptaddress]{revtex4-2}
\usepackage{graphicx}
\usepackage{epstopdf}

\usepackage[english]{babel}

\begin{document}

\title{Transformation of an energy spectrum and wave functions in the crossover from two- to three-dimensional topological insulator in HgTe quantum wells: long and thorny way}

\author{G. M.~Minkov}

\affiliation{School of Natural Sciences and Mathematics, Ural Federal University,
620002 Ekaterinburg, Russia}

\affiliation{M.~N.~Miheev Institute of Metal Physics of Ural Branch of
Russian Academy of Sciences, 620137 Ekaterinburg, Russia}

\author{V.\,Ya.~Aleshkin}
\affiliation{Institute for Physics of Microstructures  RAS, 603087 Nizhny Novgorod, Russia}

\affiliation{Lobachevsky University of Nizhny Novgorod, 603950
Nizhny Novgorod, Russia}

\author{O.\,E.~Rut}
\affiliation{School of Natural Sciences and Mathematics, Ural Federal University,
620002 Ekaterinburg, Russia}

\author{A.\,A.~Sherstobitov}

\affiliation{School of Natural Sciences and Mathematics, Ural Federal University,
620002 Ekaterinburg, Russia}

\affiliation{M.~N.~Mikheev Institute of Metal Physics of Ural Branch of
Russian Academy of Sciences, 620137 Ekaterinburg, Russia}

\author{S.\,A.~Dvoretski}

\affiliation{Institute of Semiconductor Physics RAS, 630090
Novosibirsk, Russia}

\author{N.\,N.~Mikhailov}

\affiliation{Institute of Semiconductor Physics RAS, 630090
Novosibirsk, Russia}
\affiliation{Novosibirsk State University, Novosibirsk 630090, Russia}

\author{A.~V.~Germanenko}

\affiliation{School of Natural Sciences and Mathematics, Ural Federal University,
620002 Ekaterinburg, Russia}

\date{\today}

\begin{abstract}
A magnetotransport and quantum capacitance of the two-dimensional electron gas in HgTe/Cd$_x$Hg$_{1-x}$Te quantum wells of a width  ($20.2-46.0$)~nm are experimentally investigated.  It is shown that  the first energy subband of spatial quantization is split  due to the spin-orbit interaction and the split branches are single-spin,  therewith the splitting strength increases with the increase of the quantum well width. The electron effective masses in the branches are close to each other  within the actual density range. Magneto-intersubband oscillations (MISO) observed in the structures under study exhibit the growing amplitude with the increasing electron density that contradicts to the expected decrease of wave function overlap for the rectangular quantum well. To interpret the data obtained, we have used a self-consistent approach to calculate the electron energy spectrum and the wave function within framework of the \emph{kP}-model. It has been in particular shown that the MISO amplitude increase results from the increasing overlap of the wave functions  due to their shift from the gate electrode with the gate voltage increase known as phenomenon of the negative electron polarizability. The results obtained from the transport experiments are supported by quantum capacitance measurements.
\end{abstract}

\pacs{73.20.Fz, 73.21.Fg, 73.63.Hs}

\maketitle

\section{Introduction}
\label{sec:intr}
Structures with HgTe quantum wells (QWs) are attracting a lot of attention for many reasons. First, the QW is a gapless semiconductor, while the barriers Hg$_{1-x}$Cd${_x}$Te are a semiconductor with a normal band ordering \cite{Groves63,Minkov20}.

Second, the band spectra of the parent materials HgTe and Hg$_{1-x}$Cd${_x}$Te have been studied in detail and their parameters are fairly well known.

Third, the electron energy spectrum  of the HgTe-based quantum wells is calculated within framework of the \emph{kP} theory in numerous papers (see, e.g., \cite{Gerchikov90,Zhang01,Novik05,Bernevig06,ZholudevPhD,Ren2016} and references therein). It is shown that different types of spectrum are realized depending on the QW width ($d$).  At $d = d_c\simeq  6.3$ nm it is gapless \cite{Gerchikov90} and close to the linear Dirac-like spectrum at small quasimomentum \cite{Bernevig06}. When  QW is narrow, $d < d_c$, the ordering of energy subbands of spatial quantization is analogous to that in conventional semiconductors; the highest valence subband at $k = 0$ is
formed from the heavy hole $\Gamma_8$ states, while the lowest conduction
subband is formed both from the $\Gamma_6$ states and light hole $\Gamma_8$
states. For a thicker HgTe layer, $d > d_c$, the quantum well is in
the inverted regime; the lowest conduction subband is formed from
the heavy hole $\Gamma_8$ states \cite{Dyak82e}, whereas the subband formed
from the $\Gamma_6$ states and light hole $\Gamma_8$ states sinks into the
valence band. In the inverted regime,  the spectrum becomes semimetallic when $d\gtrsim 15$~nm.

Fourth, the theory predicts that at $d> 6.5$~nm, the QW will be a two-dimensional topological insulator when, along with two-dimensional states, one-dimensional edge states are formed. And when $d$ is greater than $(60-80)$~nm it will be a three-dimensional topological insulator, when, along with two-dimensional states, two-dimensional, single-spin surface states are formed with a characteristic localization length $L_z$ in the $z$ direction less than the QW width, where $z$ is direction perpendicular to the QW plane.

And finally, when $d$ is much larger than $80$~nm, we obtain HgTe film with three-dimensional states and two-dimensional single-spin surface states.

In addition, the technology of growing HgTe/Hg$_{1-x}$Cd${_x}$Te structures is well developed \cite{Becker03,Mikhailov06,Dvoretsky10} which makes it possible to grow structures with high mobility electrons and holes. Multiband $kP$-calculations of the spectrum and wave functions are well developed and tested. This would seem to make it possible to understand in all the properties (transport, optical, etc.) of such  structures.

Experimentally, the energy spectrum of structures with $d = (4-20)$~nm detail has been studied in sufficient detail by various techniques including the optical and photoelectric methods in a wide range of radiation wavelength, starting from the terahertz range \cite{Shuvaev12,Zholudev12,Zoth14,Dantscher15}, magnetotransport \cite{Gusev12,Minkov13,Kozlov14,Minkov14,Jost17}. It was shown that in general the energy spectrum is reasonably described within the framework of the  \emph{kP} model, in which the spin-orbit (SO) interaction is taken into account.  The main contribution to SO interaction comes from the asymmetry of the interfaces for the  valence band \cite{Minkov17,Minkov20-1} and  from the Bychkov-Rashba effect for the conduction band \cite{Zhang01,Minkov19-1}. But some discrepancies remain: the effective mass of electrons at the bottom of the conduction band at thicknesses $d = (10 - 20)$~nm turns out to be $1.5 - 2$ times less than the theoretical one (the authors of Ref.~\cite{Minkov20} assume that this is due to many-particle effects).

Many papers have been devoted to the study of conductivity by the edge states \cite{Koenig07,Roth09,Gusev11,Minkov15-1}, which should be topologically protected from back scattering according to theoretical predictions \cite{Moore07,Hasan11,Qi11,Olshanetsky15}. However, in most experiments, the mean free path for such states did not exceed $1 - 3$~$\mu$m, which is not much greater than the mean free path of two-dimensional electrons ($L_p$) in these structures [for 2D systems with $\mu=(10^5-10^6)$~cm$^2$/Vs, $n=1\times 10^{11}$~cm$^{-2}$, $L_p=(0.5-5)$~$\mu$m]. Another prediction of the theory concerns the features of the spectrum and wave functions in wide HgTe quantum wells. It was predicted \cite{Zhang13} that ``single-spin'' surface states should exist in sufficiently wide quantum wells, which are  well localized at the boundaries of the quantum well. Studies of such structures have attracted a lot of attention of experimenters \cite{Bruene14,Kozlov14,Kozlov15,Savchenko16}. Structures with a QW width of $(60-90)$~nm were mainly studied.  Only the phenomenological model of surface states was used to interpret the data. No quantitative comparison with theoretical calculations was made.

What are the specific features of the methods used to study the spectrum, its transformation with increasing width, the properties of states caused by a nontrivial topology: ``edge'' and ``surface''?

In interband optics and magneto-optics experiments, both the valence band and the conduction band participate in the  processes. But the valence band has a complex spectrum, which makes it difficult to interpret experimental data.

When studying the low-temperature transport, one  studies the spectrum at the Fermi level. The Fermi energy can be easily changed, if you are investigating gated structures by changing the gate voltage ($V_g$). But in this case, the gate voltage changes not only the carrier density, and hence the Fermi energy, but also the potential profile of QW. To interpret the results in this case, it is necessary to solve the self-consistent problem taking into account the real potential, which changes significantly with a change of the gate voltage.

In principle, the Fermi energy can be changed by doping, but even in this case, it is imperative to take into account the electrostatic potential of the charge carriers and the doping impurities, i.e., to solve a self-consistent problem when interpreting the data quantitatively. In addition, these will be different samples, the asymmetry of which is difficult to determine and change.

In almost all the experimental studies of the  magnetotransport in wide QWs ($d>15 - 20$~nm), structures with the gate were studied, and the results were analyzed using an ``intuitive model'' in which it is assumed that near each of the walls there are single-spin states at a some distance from the wall of a some width $L_z$, which are independent of $V_g$ and   shift only in the energy by applied gate voltage \cite{Kozlov16,Jost17,Ziegler20}. The accuracy of this approach, especially for sufficiently wide QWs, remains unclear.

\begin{table*}
\caption{The parameters of  heterostructures under study}
\label{tab1}
\begin{ruledtabular}
\begin{tabular}{cccccccc}
structure & $d$ (nm)&  $n,p$(cm$^{-2})^\text{a}$ & $\mu_e$~(cm$^2$/Vs)$^\text{b}$& $n_{sym}$(cm$^{-2}$)& $n^{(2)}$ (cm$^{-2}$)\\
\colrule
 110614 & $20.2$ & $p=0.8\times10^{11}$   &$70000$  & $2.0\times10^{11}$ &   $>7\times 10^{11}$ \\
 180820 & $22.0$ & $n=4.6\times10^{11}$   &$265000$  &  $3.0\times10^{11}$ &  $6.3\times 10^{11}$  \\
 180824 & $32.0$ & $p=0.25\times10^{11}$   &$560000$  &  $1.5\times10^{11}$ &  $3.6\times 10^{11}$  \\
 180823 & $46.0$ & $p=0.6\times10^{11}$   & $490000$ &    $0.9\times10^{11}$ & $2.9\times 10^{11}$ \\
\end{tabular}
\end{ruledtabular}
\footnotetext[1]{For $V_g=0$~V}
\footnotetext[2]{For $n=2\times10^{11}$~cm$^{-2}$}
\end{table*}

To understand how the energy spectrum and wave functions of states change with an increase in the QW width, in this work we have experimentally studied the magnetotransport phenomena and the dependence of the capacitance between 2D gas and the gate electrode in a wide range of gate voltages in practically unexplored QWs of  ($20.2-46.0$)~nm width. When analyzing the results, we used self-consistent calculations of the energy spectrum carried out within the framework of the four-band \emph{kP} model.

\section{The structures investigated}

Our HgTe quantum wells were realized on the basis of HgTe/Hg$_{1-x}$Cd$_x$Te
($x=0.5 - 0.7$) heterostructure grown by molecular beam epitaxy on GaAs substrate with the (013) surface orientation \cite{Mikhailov06}. The nominal widths of the quantum wells under study were $d=(20.2 - 46.0)$~nm. The samples were mesa etched into standard Hall bars of $0.5$~mm width with the distance between the potential probes of $0.5$~mm. To change and control the electron and hole densities ($n$ and $p$, respectively) in the quantum well, the field-effect transistors were fabricated with parylene as an insulator and aluminum as a gate electrode. The measurements were performed in the DC  regime of linear response  at temperatures $(3 - 20)$~K  in the magnetic field up to $2.5$~T. For each heterostructure, several samples were fabricated and studied.  The parameters of the structures investigated are presented in the Table~\ref{tab1}.

The main results for all the structures investigated are qualitatively close to each other, therefore, as an example, let us consider in more detail the data obtained for the structure with $d=32$~nm, for which all the features manifest themselves more clearly.

\begin{figure}
\includegraphics[width=1.0\linewidth,clip=true]{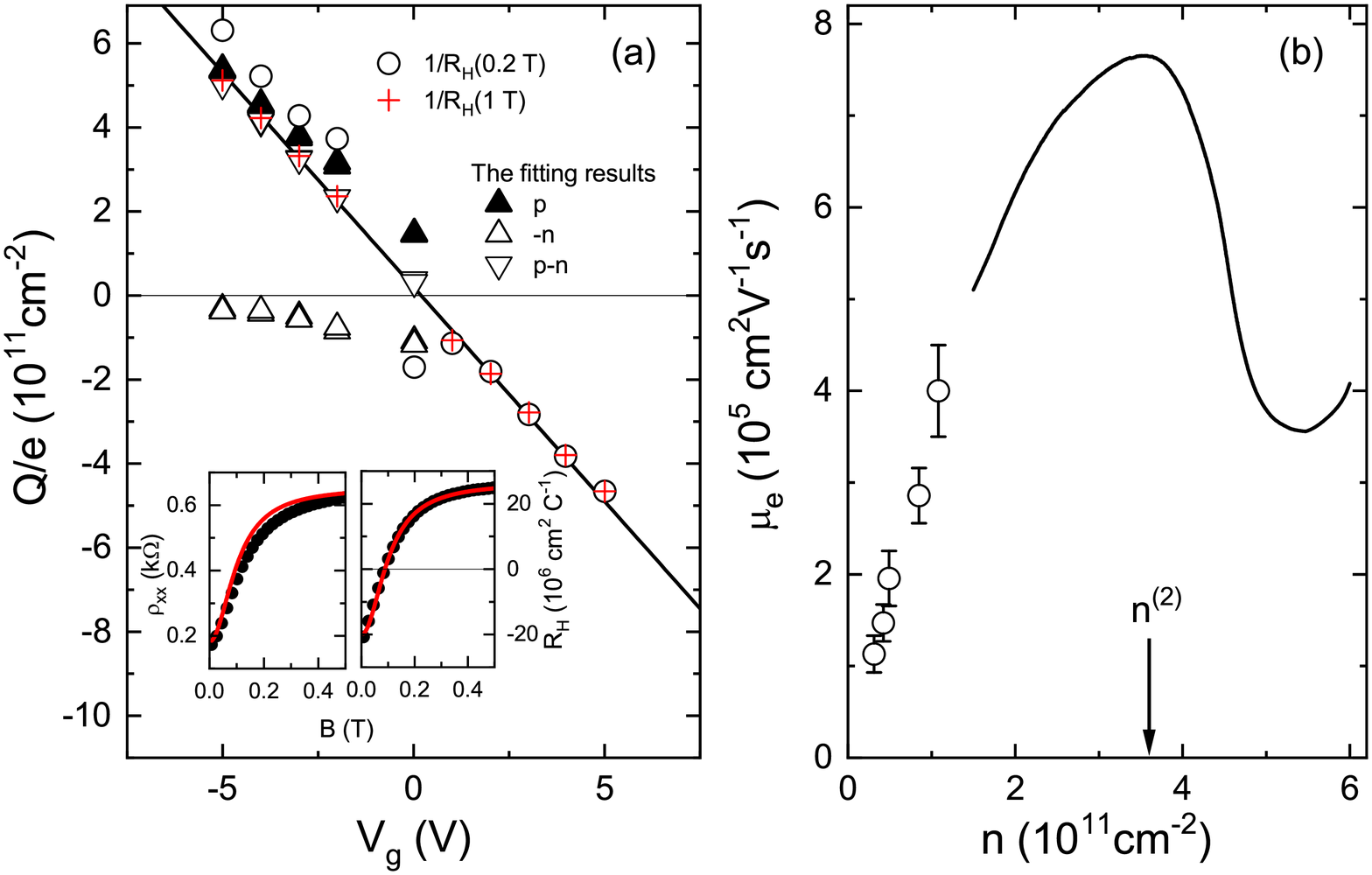}
\caption{(Color online) (a) The gate voltage dependences of electron and hole charges obtained as $1/R_H(0.2\text{ T})$ ($\circ$),   $1/R_H(1.0\text{ T})$ ($+$) and found from the fit of the data within framework of the two-types carriers model ($\vartriangle$, $\blacktriangle$). The straight line is  the dependence  $Q/e=(0.23-1.05\,V_g)\times 10^{11}$,~cm$^{-2}$. The inset  shows an example of the simultaneous fit of dependences  $\rho_{xx}(B)$ and $R_H(B)$ at $V_g = -2$~V within framework of the two-types carriers model. The symbols are the experiment, the  curves are the fitting results. (b) The $V_g$ dependence of the electron mobility found from the fit of the experimental dependences $\rho_{xx}(B)$ and $R_H(B)$ (the circles) and obtained  as   $-R_H/\rho_{xx}$ at $B=0.2$~T when $V_g$ is swept (the curve). $T=4$~K. Structure with $d=32.0$~nm.}
\label{f1}
\end{figure}

For a general characterization of the structure, let us consider first the gate voltage dependences of the electrons and holes over entire range of $V_g$. When $V_g > 1$~V, only electrons participate in the conductivity, the Hall resistance $\rho_{xy}$ at low magnetic field ($B<0.2 - 0.4$~T) linearly depends on $B$, and the Hall electron density $n = - 1/eR_H$, where $e$ and $R_H$ stand for elementary charge and the Hall coefficient, respectively,  increases linearly with the $V_g$ increase [Fig.~\ref{f1}(a)]. When $V_g <1$~V, the Hall coefficient strongly depends on $B$, changing sign from electronic at low $B$ to hole-like with increasing $B$, and at the same time $\rho_{xx}$ increases strongly with $B$ [see insets in Fig.~\ref{f1}(a)]. Within this gate voltage range the electron and hole densities were found by simultaneous fit of dependences $\rho_{xx}(B)$ and $R_H(B)$  at $B<0.3$~T within classical model of the conductivity by two types of the carriers. Fig.~\ref{f1}(a) shows that total charge of free carriers in QW $Q=e(p-n)$ linearly depends on $V_g$ as  $Q/e=(0.23-1.05 V_g)\times 10^{11}$,~cm$^{-2}$ within whole $V_g$ range. Note that the absolute value of the slope of this dependence $1.05\times 10^{11}$~cm$^{-2}$V$^{-1}$  is in a good agreement with that obtained from the capacitance measurements $edn/dV_g=C/S_g$, where $C$ is the capacitance between the 2D gas and the gate electrode, measured for the same structure, $S_g$ is the gated area.

Fig.~\ref{f1}(b) shows the electron density dependencies of electron mobility at $V_g<1$~V when electrons exist together with hole (circles) and at $V_g>1$~V when only electrons contributes to the conductivity (the curve). It seen that the $n$ dependence of the mobility is non-monotonic. It increases with $n$ increase while $n\lesssim 3.6\times 10^{11}$~cm$^{-2}$ and shows a sharp decrease at higher values of $n$ resulting from the beginning of filling the second subband of spatial quantization [see Table~\ref{tab1} and Fig.~\ref{f4}(b)].

\section{Spin-orbit splitting of conduction band. Experiment}

The aim of this paper is to study the spectrum and wave functions of electrons  in the conduction band, therefore, below we will discuss the results only for $V_g> 0$. The spectrum of the valence band is much more complicated and should be discussed in a separate paper.

\begin{figure}
\includegraphics[width=1.0\linewidth,clip=true]{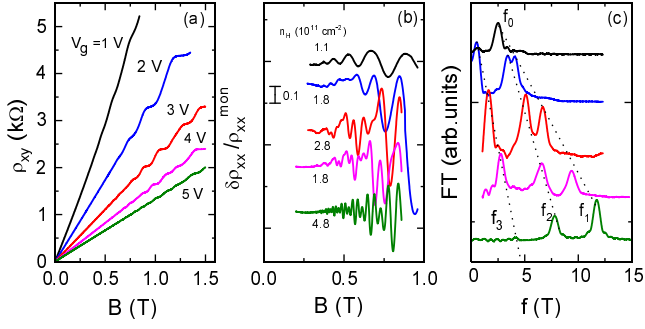}
\caption{(Color online) (a) and (b) The magnetic field dependences of $\rho_{xy}$ and oscillating part of  $\rho_{xx}$, respectively. (c) The Fourier spectrum of
of the oscillations shown in the panel (b) performed over the magnetic filed range $(0.2 - 0.6)$~T. Structure 180824 with $d=32.0$~nm. }
\label{f2}
\end{figure}

As example, the magnetic field dependences of $\rho_{xy}$ and oscillating part of resistivity $\Delta\rho_{xx}(B)=\rho_{xx}(B)-\rho^{mon}_{xx}(B)$, where $\rho^{mon}_{xx}(B)$ is the monotonic part of $\rho_{xx}(B)$,   for some gate voltages are presented in Fig.~\ref{f2}(a) and \ref{f2}(b), respectively. As seen, $\rho_ {xy}$ linearly increases with $B$ in low magnetic fields, then the  oscillations appear, which are transformed to the  steps of the quantum Hall effect  in the higher magnetic field.

The Fourier spectra of  $\rho_{xx}$ found in magnetic field range before onset of the steps of the quantum Hall effect are presented  in Fig.~\ref{f2}(c). These data show that unsplit oscillations with one component in the Fourier spectrum  $f_0$ is observed for low electron density $n\lesssim (1.4-1.5)\times 10^{11}$ cm$^{-2}$. With increasing $V_g$, it splits into two $f_1$ and  $f_2$ components, and a low-frequency $f_3$ component  appears.

\begin{figure}
\includegraphics[width=1.0\linewidth,clip=true]{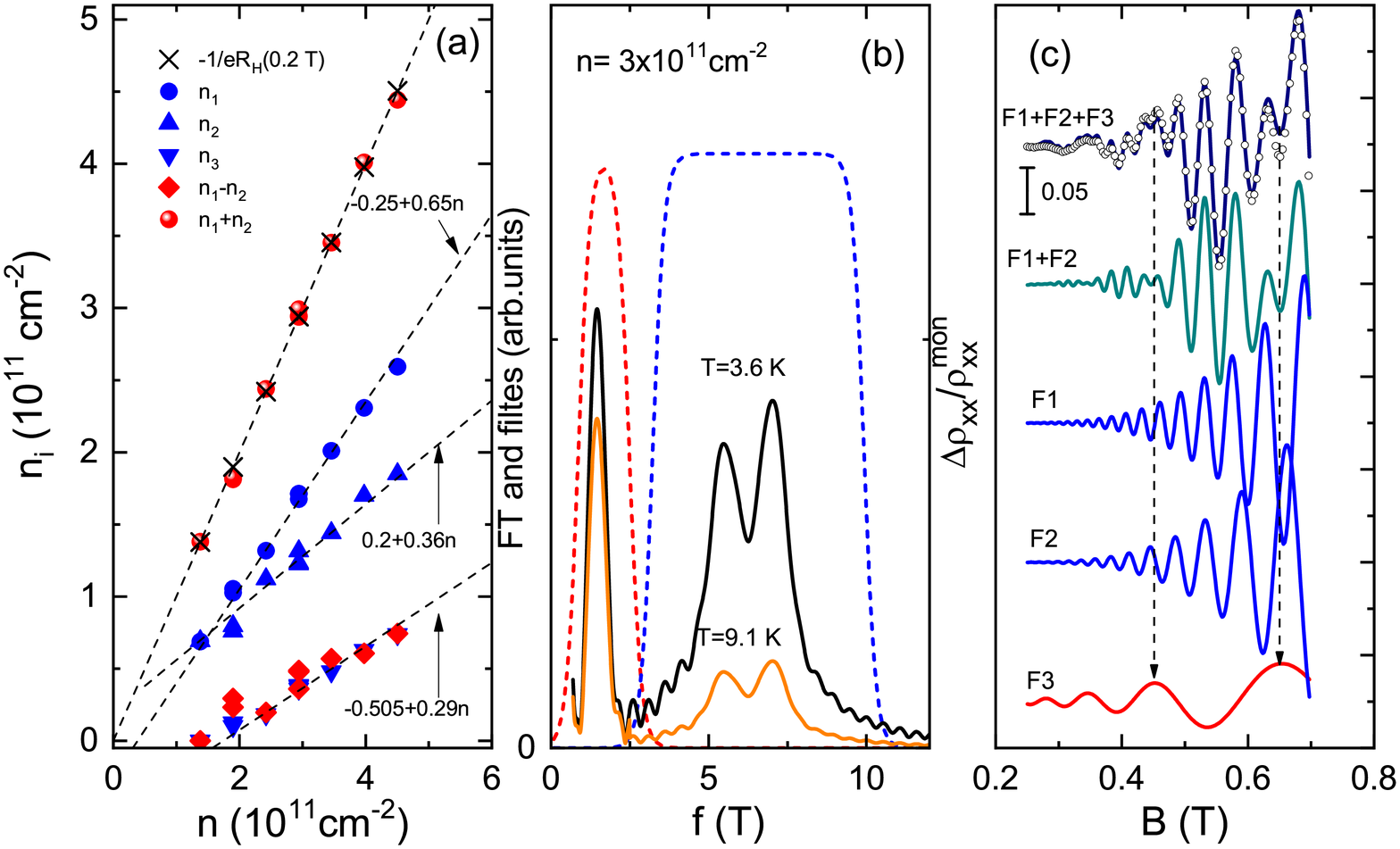}
\caption{(Color online) (a) The density of electrons in the branches, their sum and difference plotted against the total electron density, found as $n=-1/eR_H(0.2\text{ T})$. The straight lines show the results of the linear fit. $T=3.6$~K.    (b) The Fourier spectra of the oscillations at $n=3\times 10^{11}$~cm$^{-2}$ for $T=3.6$~K and $9.1$~K. The dotted curves show the characteristics of the filters used to separate the low  and high-frequencies components. (c)  The oscillating part of resistivity. The points are the experimental dependence. The curves are the results of data analysis (see the text). $T=3.6$~K. Structure with $d=32.0$~nm. }
\label{f3}
\end{figure}

At $n> (1.4-1.5)\times 10^{11}$ cm$^{-2}$, the electron densities corresponding to each Fourier component are plotted in Fig.~\ref{f3}(a) assuming that these states are non-degenerate. It can be seen that the sum of the two high-frequency components $f_1 + f_2$ gives a density that coincides with the Hall density. The difference $f_1-f_2$ coincides with the low-frequency component $f_3$. This is consistent with  the following interpretation: (i) the high frequency components $f_1$ and $f_2$ corresponds to two  single-spin branches of the first spatially quantized subband split due to  SO interaction, the splitting value increases with increasing gate voltage; (ii) the low frequency component $f_3$ arises from the transitions between split branches  which result in the well-known magneto-intersubband oscillations (MISO). This interpretation is confirmed by the temperature dependence of the amplitudes of these Fourier components shown in Fig.~\ref{f3}(b). It is seen that the amplitude of the two high-frequency components corresponding to the Shubnikov-de Haas (SdH) oscillations of each of the spectrum branches decreases strongly with increasing temperature, while the MISO amplitude decreases insignificantly, as predicted  theoretically \cite{Mamani09,Raichev10}.

Figure~\ref{f3}(a) shows that the electron densities in the split branches within experimental uncertainty are well described by the linear dependences with the different slopes equal to  $sl_1=0.65$ and $sl_2=0.36$. These lines intersect at the point  $n\simeq 1.5\times 10^{11}$~cm$^{-2}$. Therefore, we believe that the quantum well is close to symmetric one at $n_{sym}\simeq 1.5\times 10^{11}$~cm$^{-2}$.

\begin{figure}
\includegraphics[width=0.7\linewidth,clip=true]{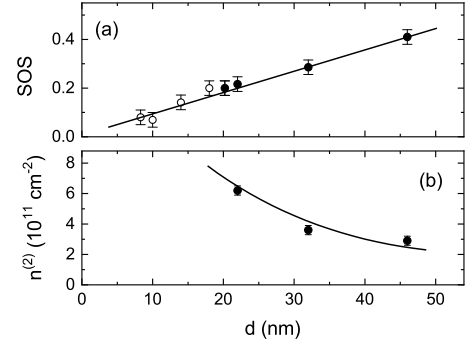}
\caption{(Color online) (a) The SOS value  for structures of different width. Solid symbols are the results of this paper, open symbols are the results obtained in Ref.~\cite{Minkov19-1}. The  curve is a guide for the eye.  (b) The density of electrons when the second sub-band begins to be occupied plotted against the QW width. The symbols are the data, the  curve is a guide for the eye.}
\label{f4}
\end{figure}

To compare the results obtained for the  structures with the different QW widths, we will characterize the ``strength'' of spin-orbit splitting  by the phenomenological parameter $SOS = (sl_1-sl_2) / (sl_1 + sl_2)$ \footnote{This is of course an approximation.  In the paper \cite{Minkov19-1} we show that the theory predicts nonlinear dependences $n_1(V_g)$ and $n_2(V_g)$   at the small splitting values. However,  in order to trace the change in spin-orbit splitting with the $d$ change, such a SOS parameter is suitable.}.
In Fig.~\ref{f4}(a), we plot the $d$ dependences of SOS for all the structures from the Table~\ref{tab1} and for the structures investigated in Ref.~\cite{Minkov19-1}.  As seen the  SOS value monotonically, without inflection, increases with the increasing QW width  over the entire $d$ range. This shows that no new strongly localized ``surface'' states arise up up to  $d=46$~nm at least.

Another parameter that can be extracted from the analysis of the SdH oscillations is the electron density $n^{(2)}$ at which the second subband of the spatial quantization begins to be occupied. As seen from Fig.~\ref{f4}(b) the $n^{(2)}$ values diminish with $d$ increase  monotonically that as shown in Section~\ref{sec:cr} agrees satisfactorily with the theoretical results.

Thus, all the results presented above show that the main magnetotransport properties of  wide QWs are qualitatively similar to that previously investigated in QWs with a width of $d <20$~nm.

\section{Electron masses in the branches}
\label{sec:mass}

To to separate the SdH oscillation components corresponding to split branches and obtain the masses in them, we used of bandpass filtering as shown in Fig.~\ref{f3}(b). To improve the resolution of the Fourier spectra, the oscillating part of $\rho_{xx}(B)$ was multiplied by $1/B$. Then,  applying the inverse Fourier transformation we obtained  the oscillations of the   frequency $f_3$ [the curve F3 in Fig.~\ref{f3}(c)] and superposition of oscillations with two higher frequencies $f_1$ and $f_2$  [the curve labeled as F1+F2 in Fig.~\ref{f3}(c)]. To keep the correct ratio of the amplitudes of the different oscillations, the curves obtained after the inverse Fourier transform were multiplied by $B$. Fitting  the curve F1+F2 by the sum of the two Lifshits-Kosevich (LK) formulas  \cite{LifKos55} allows us to separate the contributions which are presented by the curves F1 and F2 in this figure (for more detail see Ref.~\cite{Minkov20-2}). This method of decomposition allows us not only to reliably separate out the contributions of each of the three oscillation components and determine their frequencies, but also to determine the amplitudes. Applying this procedure to analyse the oscillations measured for different temperatures one can be obtain the temperature dependences of the amplitudes of the oscillations coming from the branches  and hence obtain the cyclotron effective mass $m=\hbar^2k(dE/dk)^{-1}$ at the Fermi energy for each of these branches.
\begin{figure}
\includegraphics[width=0.7\linewidth,clip=true]{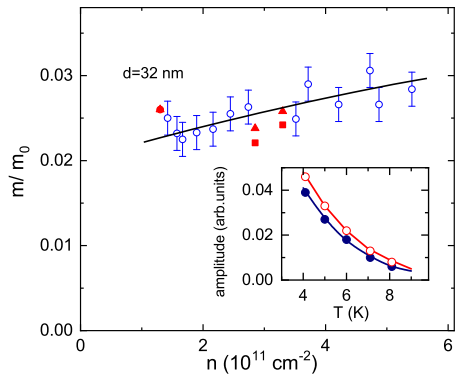}
\caption{(Color online) The electron density dependence of the electron effective mass at the Fermi energy for the structure with $d=32.0$~nm.   The triangles and squares are obtained from the temperature dependences of the amplitudes of the separated  SdH oscillations $\rho_{xx}(B)$ and correspond to the masses in each of the branches of the spectrum. The circles are obtained from the temperature dependence of the oscillation amplitude of $d^2\rho_{xx}/dV_g^2$, measured with a change in the gate voltage at $B = 0.5$~T and corresponds to the average value of the masses in the branches.  The line is a guide for the eye. The inset  shows the temperature dependences of the amplitudes of separated components of the SdH oscillations for $n=3.3\times 10^{11}$~cm$^{-2}$ at $B=0.5$~T (symbols) and the results of the best fit to the LK formula (lines). }
\label{f5}
\end{figure}

The inset in Fig.~\ref{f5}  shows the temperature dependences of the amplitudes of the SdH oscillation components for $B=0.5$~T separated as described above for the total electron density $n=3.3\times 10^{11}$~cm$^{-2}$. It is seen that the temperature dependences of the amplitudes are very close to each other. They are well fitted by the LK formula that gives the values of effective masses $m   = (0.0258\pm 0.0015)m_0$ and $(0.0242\pm 0.0015)m_0$ for the electron densities in the branches equal to $1.9\times 10^{11}$~cm$^{-2}$ and $1.42\times 10^{11}$~cm$^{-2}$, respectively. These values and the values obtained for the other electron densities are shown in Fig.~\ref{f5} by squares and triangles.  As seen the effective masses in the spin-orbit splitting branches are very close to each other.

\begin{figure}
\includegraphics[width=0.8\linewidth,clip=true]{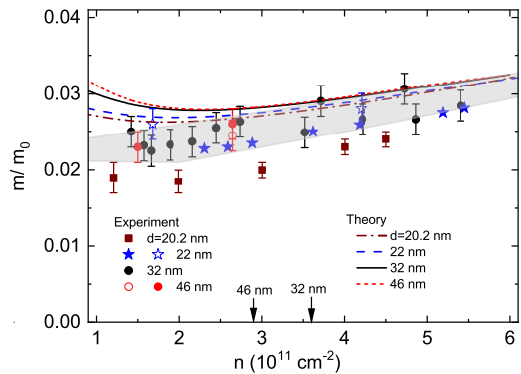}
\caption{(Color online) The electron density dependence of the electron effective mass for the structures with different QW width. The symbols are the data,  the lines are the calculated dependences for rectangular quantum well. The shadow area is the area within which the experimental data for QWs with $d=(22.0-46.0)$~nm are. The arrows show the $n_{(2)}$ values for $d=32$~nm and $46$~nm.}
\label{f6}
\end{figure}

We attract here the reader's attention  to a peculiarity of MISO in HgTe-based QWs already discussed in Ref.~\cite{Minkov20-2} for the narrower quantum wells.  Fig.~\ref{f3}(c) shows that beating of the SdH oscillations results as a sum of two high-frequency components. As seen the magnetic fields of the antinodes of high-frequency oscillations correspond to the minima in $\rho_{xx}$-MISO. Such a mutual position is opposite to that observed in ordinary structures, namely, in double  and wide quantum wells. But it agrees with that observed in narrow HgTe- and In$_{x}$Ga$_{1-x}$As-based QWs, in which the splitting of oscillation arises due to spin-orbit splitting. The authors of Ref.~\cite{Minkov20-2} assume
that the unusual mutual positions of the MISO extrema and the SdH oscillation antinodes originate from the dependence of the probability of transitions between the Landau levels of different branches on the difference in their energies.

Let us consider the data obtained by the  other method which gives the effective mass over a wide electron density range. It consists in  analysing  the temperature dependence of the  amplitude of the oscillations of the $d^2\rho_{xx} / dV_g^2$ vs $V_g$ dependences in a fixed magnetic field.  This method gives an average of the effective masses in the branches.  The results are shown in Fig.~\ref{f5} by circles. It is seen that the results obtained by the two methods practically coincide. The effective masses increase slightly with an increase in the total electron concentration from $(0.020 \pm 0.003)m_0$ at $n=1.5\times 10^{11}$~cm$^{-2}$ to $(0.030 \pm 0.003)m_0$ at $n=4.5\times 10^{11}$~cm$^{-2}$.

The same measurements and analyses were carried out for all the structures under study. The results  are shown in Fig.~\ref{f6}. It is seen that effective masses QWs with $d=(22.0-46.0)$~nm   are close to each other  over the whole electron density range. They are within the shadow area which width  is comparable with the experimental errors. In the same figure, we plot the theoretical dependences calculated within the framework of the standard \emph{kP} model for the rectangular quantum well when the potential of  electric charge is not taken into account. It is seen that the theoretical dependences and experimental plot are qualitatively similar. As for the data point for the QW with $d=20.2$~nm, the low values of the effective mass in the narrow QWs, $d = (7-20)$~nm,  is already observed and discussed earlier \cite{Minkov20}.

Let us now consider the behavior of MISO with changing the electron density in QWs of different widths.

\section{MISO amplitude with changing electron density and QW width}
\label{sec:miso}

\begin{figure}
\includegraphics[width=1.0\linewidth,clip=true]{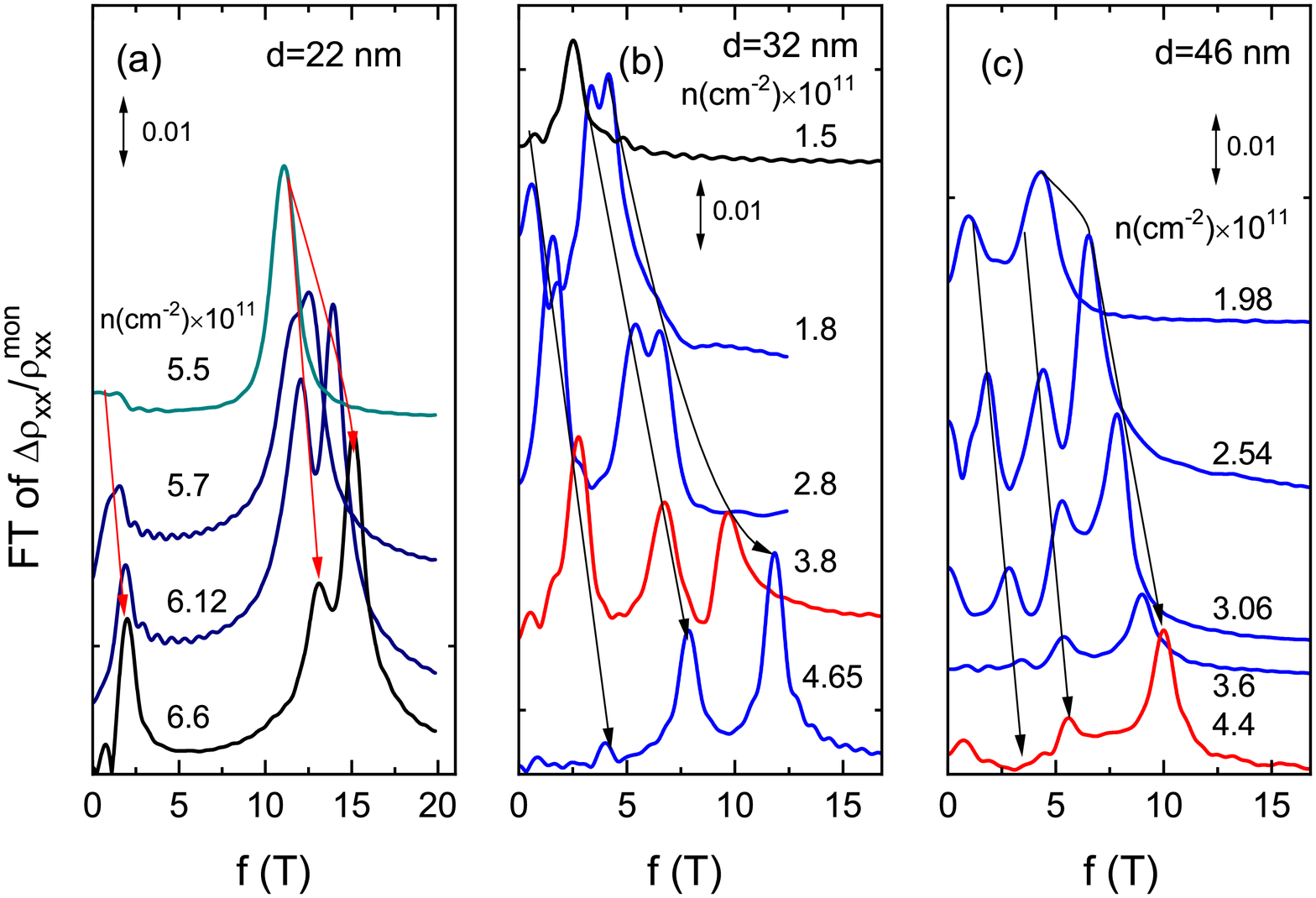}
\caption{(Color online)  The Fourier spectra of $\Delta \rho_{xx}/\rho_{xx}^{mon}$ for structures investigated. The Fourier transformation was performed without correcting the dependence of the amplitude of oscillations on the magnetic field as has been done for  Fig.~\ref{f3}, t. For clarity, the curves are shifted along the vertical axis. $T=3.6$~K. }
\label{f7}
\end{figure}

The Fourier spectra of the oscillating part of $\rho_{xx}(B)$ for three  structures for some electron densities are presented in Fig.~\ref{f7} (in contrast to Fig.~\ref{f3}, the Fourier transformation was performed here without correcting the dependence of the amplitude of oscillations on the magnetic field). Let us first compare the strength of  MISO and  SdH oscillations qualitatively. At first glance, the ratio of the amplitude of the Fourier peaks of  MISO and SdH oscillations is approximately the same in structures with different QW widths. Seemingly it contradicts to intuitive expectations and theoretical predictions  according to which the overlapping of the wave functions of two branches should decrease with the increasing QW width that should result in decrease of the probability of transitions between them, and, consequently, to a decrease in the MISO amplitude.

\begin{figure}
\includegraphics[width=0.7\linewidth,clip=true]{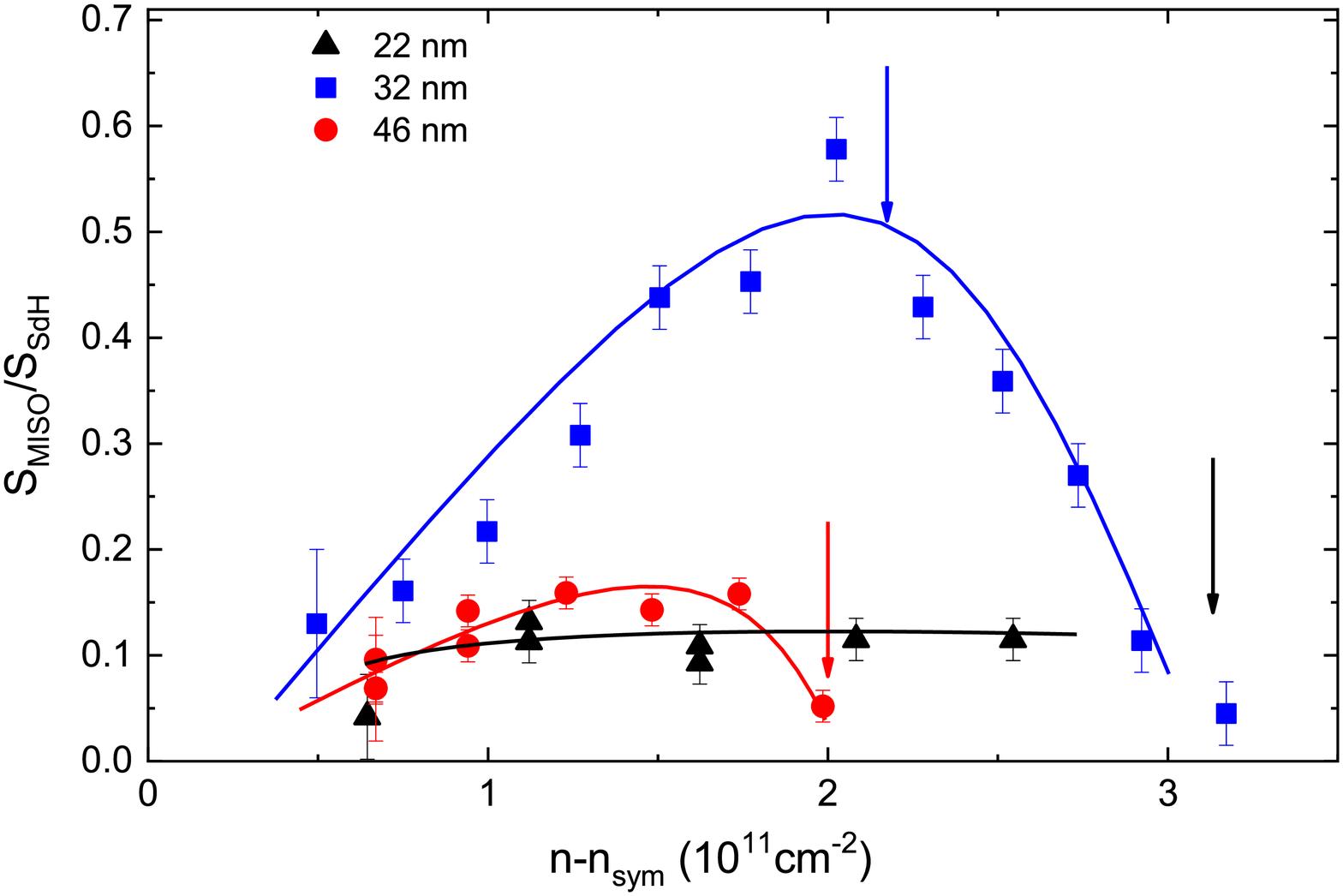}
\caption{(Color online)  The $S_\text{MISO}/S_\text{SdH}$ value plotted as a function of $n-n_{sym}$. The lines are a guide to the eye. The arrows show the densities at which the second subbands begins to be occupied.}
\label{f8}
\end{figure}

It would seem that by separating the oscillations, we find both prefactors and exponentials that describe the dependence of the oscillations on the magnetic field, and then one should compare these parameters for the SdH oscillations and MISO.   However, our analysis shows that each of these parameters is found with not enough accuracy, only their combination is obtained well. Therefore, we restrict ourselves to comparing the areas under the Fourier peaks of the spectra corresponding to MISO and SdH oscillations, denoted as $S_\text{MISO}$ and $S_\text{SdH}$, respectively. For this, in contrast to Fig.~\ref{f3}, the Fourier transformation  was performed without correcting the dependence of the amplitude of oscillations on the magnetic
field. The $S_\text{MISO}/S_\text{SdH}$ value plotted against $(n-n_{sym})$ are shown in Fig.~\ref{f8}.

It is seen that for a small splitting, that corresponds to $n-n_{sym} \lesssim 1\times 10^{11}$~cm$^{-2}$, the ratio $S_\text{MISO}/S_\text{SdH}$ does not change much with increasing $d$, although the overlap of the wave functions of different branches should seem to decrease strongly.

\begin{figure}
\includegraphics[width=0.8\linewidth,clip=true]{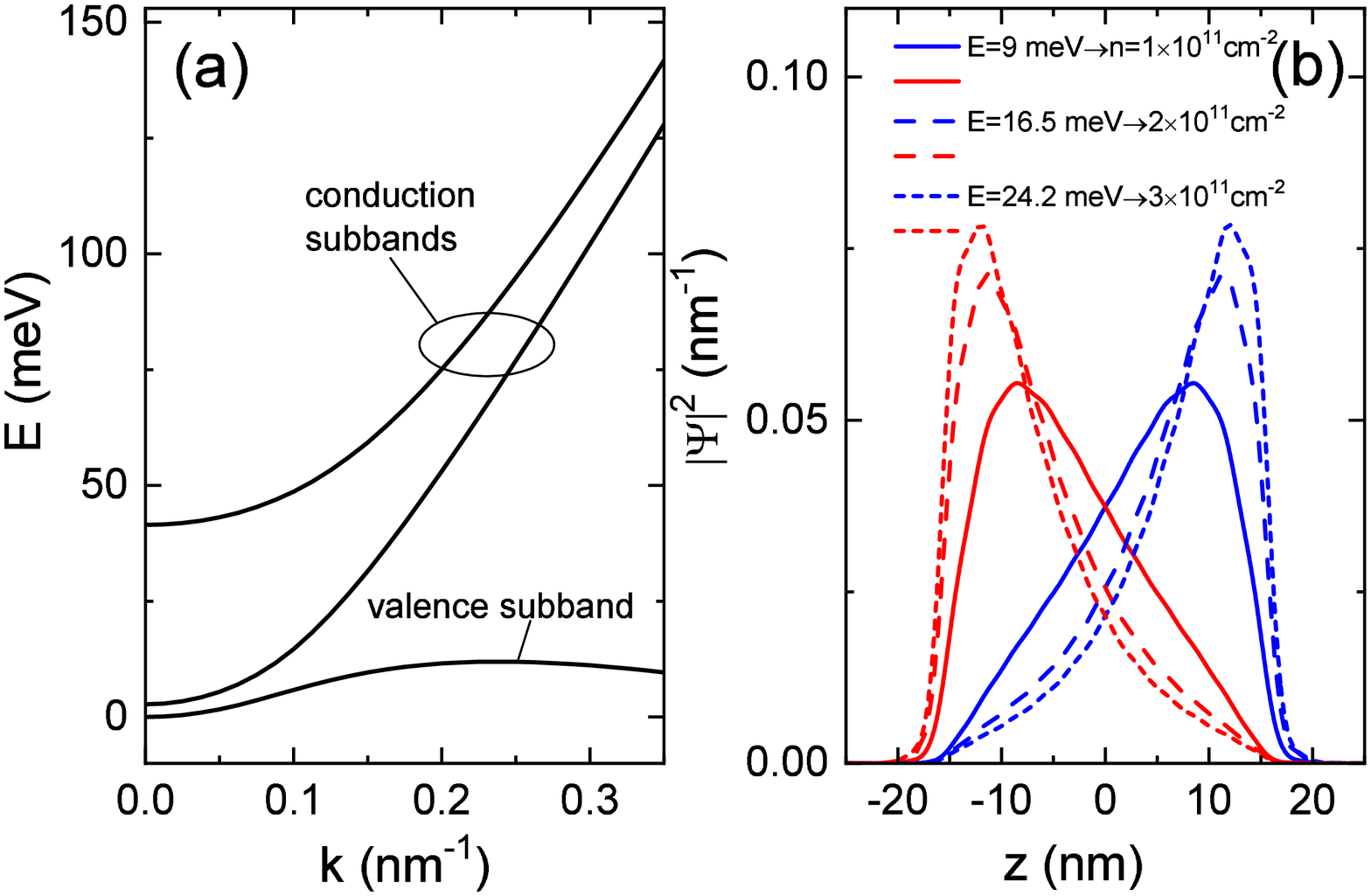}
\caption{(Color online)  (a) The energy spectrum of the rectangular  QW of $32$~nm width. (b) $|\Psi_s (z)|^2$ for two states of the conduction band at different energies corresponding to the different electron densities.}
\label{f9}
\end{figure}

Most important in Fig.~\ref{f8} in our opinion is the difference in the dependences of $S_\text{MISO}/S_\text{SdH}$ on $n-n_{sym}$. As seen  the ratio $S_\text{MISO}/S_\text{SdH}$ for QW with $d=22$~nm  does not depend on the density within the experimental accuracy\footnote{The first point [with $(n-n_{sym})=0.65\times 10^{11}$~cm$^{-2}$], when the MISO just appears, has a large error, it corresponds to the low frequencies and it is difficult to ``clear'' it from the monotonic component}. In wells with $d =32$~nm and $46~$~nm, the $S_\text{MISO}/S_\text{SdH}$ value noticeably increases with an increase of  $n-n_{sym}$. Such a behavior seems strange, because with an increase of electron density (i.e., with an increase of the Fermi quasimomentum) the overlapping of the wave functions of two branches located at different QW walls decreases at least in the empty spectrum [see Fig.~\ref{f9}(b)]. Figure~\ref{f8} also shows sharp drops in $S_\text{MISO}/S_\text{SdH}$ at $(n-n_{sym})\simeq 3\times 10^{11}$ and $1.6\times 10^{11}$~cm$^{-2}$ for structures with $d=32$~nm and $46$~nm, respectively. These electron densities are close to the beginning of the filling of the second subband of spatial quantization [see Table~\ref{tab1}, Figs.~\ref{f1}(b) and \ref{f4}(b)].

In order to understand and interpret all the results presented above, namely, the electron density dependences of the splitting magnitude, masses in the branches, MISO amplitude in the quantum wells of different widths, let us consider the predictions of the \emph{kP} theory.

\section{Results of calculation: the energy spectrum and wave functions}
\label{sec:cr}

The energy spectrum for the HgTe quantum wells was calculated using the Kane’s four-band model including second order remote band contributions. The expression for the $8\times 8$ Hamiltonian of the heterostructure grown on the plane of different orientations was derived by the method described in Refs.~\cite{Los96,Becker00}. The components of the built-in strain tensor were calculated with the use of the formulas from \cite{Novik05}. Parameters to describe the deformation contribution to the Hamiltonian were
taken from Ref.~\cite{Takita79}. To take into account the interface inversion asymmetry we used an additional term in the Hamiltonian, which is suggested by Ivchenko \cite{Ivch05} (see also Ref.~\cite{Minkov17} for more details).

To account for the influence of the electrons on the electric field distribution across the quantum well, the standard self-consistent procedure of simultaneous solution of the Shr\"{o}dinger equation and Poisson equation
\begin{eqnarray}
  \frac{d}{dz}\left(\kappa(z)\frac{d}{dz}\varphi(z)\right)&=& \frac{e}{(2\pi)^2} \label{eq10} \\ 
  &\times & \sum_s\int d^2\mathbf{k}|\Psi_s(\mathbf{k})|^2 f[\varepsilon_s(\mathbf{k})]-\frac{eN_D^+}{d} \nonumber
\end{eqnarray}
was used. Here, $\kappa(z)$ is the dielectric susceptibility, $\varphi(z)$ is the electrostatic potential, $f[\varepsilon_s(\mathbf{k})]$ is the Fermi-Dirac distribution function, $\Psi_s(\mathbf{k})$ is the wave function of eight components, which has the form
\begin{equation}\label{eq05}
  \Psi_s(\mathbf{k})=C_s\exp{\left(ik_xx+ik_yy\right)\phi_s(z,k_x,k_y)},
\end{equation}
$N_D^+$ is the density of charged donors (it is supposed that the donors are distributed uniformly across the quantum well \footnote{It should be mentioned that the results under consideration are practically insensitive to where the donors are located -- in the quantum well or in the barriers.} and their density is equal to the electron density at the gate voltage for which the quantum well is symmetric $N_D^+=n_{sym}$), the summation in Eq.~(\ref{eq10}) runs over the spatial quantization subband including the ``spin'' index. To solve the Shr\"{o}dinger equation we set the boundary conditions in the plane $z=-z_0$ assuming that the center of the quantum well was located at $z=0$. The value of $z_0$ was chosen in such a way that the wave functions at $z=\pm z_0$ were practically equal to  zero. Usually, it was enough for the value of $z_0$ to exceed the width of the quantum well by $40$~nm. Eq.~(\ref{eq10}) was solved with the following boundary conditions
\begin{eqnarray}\label{eq20}
  \varphi\left(-z_0\right)&=&0 \nonumber \\
  \left.\frac{d\varphi(z)}{dz}\right|_{z=-z_0}&=&0,
\end{eqnarray}
where is assumed that the gate electrode is located at positive  $z$.

To calculate the energy spectrum and wave functions for a given $n$ value, we solved simultaneously the Schr\"{o}dinger and Poisson equations by using an iterative method. First, the electron spectrum is calculated for zero potential. The wave functions obtained are used to calculate the electron charge distribution  corresponding to given $n$. This distribution is substituted in right-hand side of the Poisson equation, Eq.~(\ref{eq10}). Obtained $\varphi(z)$ is substituted in the Hamiltonian and then the Schr\"{o}dinger equation is solved again.  The procedure is repeated until the potential $\varphi(z)$ converges. This usually requires $5-10$ iterations.

\begin{figure}
\includegraphics[width=1.0\linewidth,clip=true]{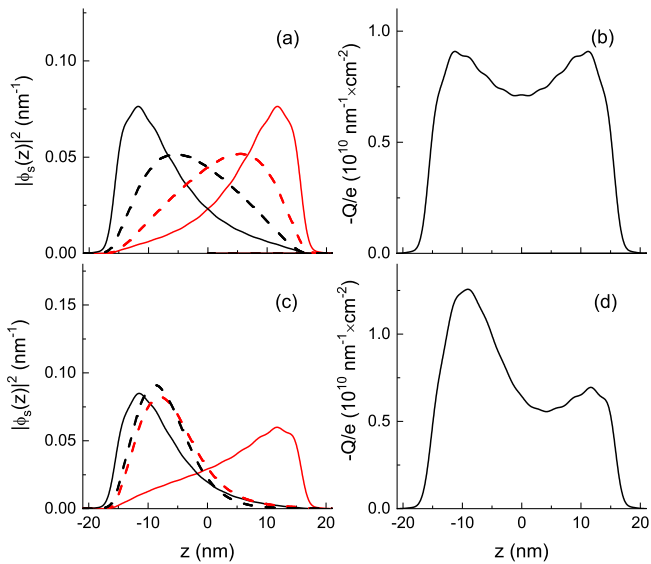}
\caption{(Color online)  (a) and (c) The dependence  $|\phi_s(z,k_F^{(s)})|^2$ for two states $s=1,2$ at $k_F^{(s)}$ (solid lines) and $0.5\,k_F^{(s)}$ (dashed lines). (c) and (d) The  distribution of electric charge. The $k_F$ value corresponds to the electron density $2.5\times 10^{11}$~cm$^{-2}$. The panels (a), (b) and (c), (d) show the calculation results obtained without and with applying the self-consistent procedure, respectively. $d=32$~nm. The gate electrode is located at $z>0$.}
\label{f10}
\end{figure}

Let us first consider the calculation results for the empty spectrum, i.e., when QW is rectangular in the shape. The dispersion $E(k)$ is depicted in Fig.~\ref{f9}(a), while the normalized sum of squares of eight components of wave functions for two states of the conduction band, $|\phi_1(z)|^2$ and  $|\phi_2(z)|^2$, for QW with $d=32$~nm are shown in Fig.~\ref{f9}(b). As seen from Fig.~\ref{f9}(b) the wave functions of each of the branches are localized near one of the  wall, and the larger $k$ value the stronger the localization and the less the overlap of the wave functions of different branches. An approach of the empty spectrum is however inapplicable to describe the experiment. This is because the charge distribution changes and, hence, the shape of the quantum well also changes when  the electron density is tuned via the  voltage on the gate electrode. Self-consistent procedure should be used for that.

An importance of the role of the  potential of electrons   in the forming shape  of the quantum well is illustrated by Fig.~\ref{f10}. The dependences $|\phi_s(z)|^2$   for $k_F^{(s)}$ (dashed lines) and at $0.5\,k_F^{(s)}$ (solid lines) for two states, $s=1,2$, for  $k_F^{(s)}$ corresponding to the electron density $2.5\times 10^{11}$~cm$^{-2}$  before self-consistency are shown in Fig.~\ref{f10}(a). The distribution of electron charge $Q(z)$ is presented in Fig.~\ref{f10}(b). As evident from the last figure  $Q(z)$ is symmetric and the average value of the electron charge position is equal to zero for this case. The same dependences after procedure of self-consistency  are shown in Figs.~\ref{f10}(c) and \ref{f10}(d). It is clearly seen that both $|\phi_s(z)|^2$ and $Q(z)$ are modified drastically.

\begin{figure}
\includegraphics[width=0.6\linewidth,clip=true]{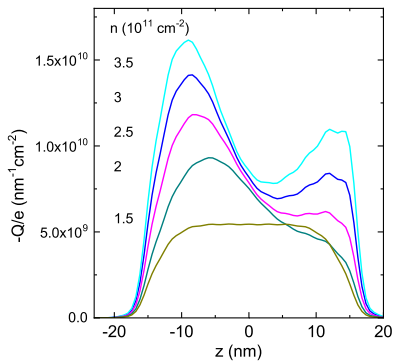}
\caption{(Color online)  $d=32$~nm. The redistribution of the charge in  QW with increasing  electron density caused by  $V_g$ increase. The gate electrode is located at $z>0$. }
\label{f11}
\end{figure}

Redistribution of the electron charge in QW with the growing gate voltage inducing the electron density increase  is shown in Fig.~\ref{f11}. As seen the charge shifts  monotonically away from the gate electrode  with an increase in the positive charge on it, first quickly, then slowly. The center of gravity of the charge distribution shifts away from the gate, which corresponds to the repulsion of the electronic state from the gate electrode charged positively. Such a behavior is contr-intuitive and corresponds to negative polarizability (this phenomenon  for analogous structures was discussed in Ref.~\cite{Aleshkin21}).

\begin{figure}
\includegraphics[width=1.0\linewidth,clip=true]{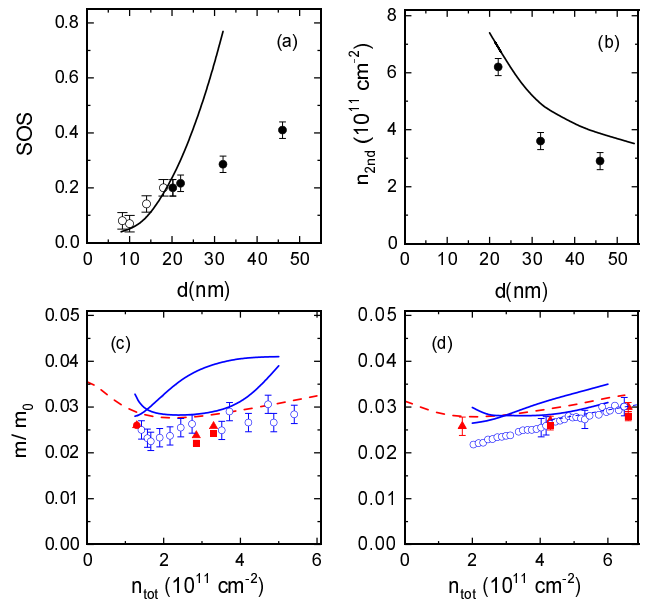}
\caption{(Color online) (a) The SOS value  for structures of different width. The symbols are the data from Fig.~\ref{f4}(a). The  curve is result of self-consistent calculation.  (b) The density of electrons when the second sub-band begins to be occupied. The symbols are the data from Fig.~\ref{f4}(b), the line is result of self-consistent calculation. (c) and (d) The electron density dependence of the electron effective mass at the Fermi energy for the structures with $d=32$~nm and $22$~nm, respectively.   The triangles and squares are obtained from the temperature dependences of the amplitudes of the separated  SdH oscillations $\rho_{xx}(B)$ and correspond to the masses in each of the branches of the spectrum. The circles are obtained from the temperature dependence of the oscillation amplitude of $d^2\rho_{xx}/dV_g^2$, measured with a change in the gate voltage at $B = 0.5$~T and corresponds to the average value of the masses in the branches.  The dashed lines are calculated for an empty well. The solid lines are the result of a self-consistent calculation of masses for each branches.  }
\label{f111}
\end{figure}

\begin{figure}
\includegraphics[width=1.0\linewidth,clip=true]{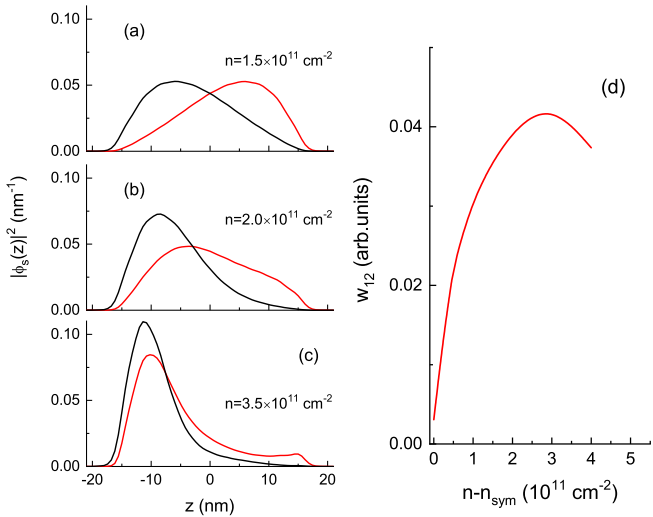}
\caption{(Color online)  (a) -- (c) The transformation of the probability density distribution  $|\phi_s(z)|^2$ at $k_F^{(s)}$ with growing electron density controlled by the gate voltage. (d) The rate of the $k_F^{(1)}\rightarrow k_F^{(2)}$ transitions due to scattering on the short range random potential. $d=32$~nm.  }
\label{f110}
\end{figure}

Now we in position to compare the results of  self-consistent calculations with our experimental data (see Fig.~\ref{f111}). Let us first consider the dependence of SOS on the QW width.   As seen from Fig.~\ref{f111}(a) both the experimental   and  calculated dependences grow with increasing $d$, and for $d <22$~nm the theoretical dependence describes the experiment well. However, at $d> (22-25)$~nm, the calculated dependence increases much faster. The calculated and experimental dependences $n^{(2)}(d)$ are in a rather good agreement [see Fig.~\ref{f111}(b)].

The comparison between  the experimental and calculated effective masses is given in Fig.~\ref{f111}(c) and Fig.~\ref{f111}(d) for $d=32$~nm and $22$~nm, respectively. As seen the calculated effective masses  are different for the split branches that qualitatively agrees with the data. However,  quantitatively, this difference is much less experimentally. The other point is that the experimental masses are less as compared with the calculated ones practically for all the electron densities. Such a discrepancy for HgTe-based QWs is already discussed earlier in Ref.~\cite{Minkov20} where assumption was made that such a behavior mey result from the many-body effects.

The theoretical model used provides qualitative understanding of the unusual behaviour of the MISO amplitude with  growing electron density considered in Section~\ref{sec:miso}. As Figs.~\ref{f110}(a)-\ref{f110}(c) illustrate the growing gate voltage results to the strengthening overlap of wave functions of the branches due to their displacement  away from the gate electrode that in its turn lead to strong increase of the rate of the transitions between different ``spin'' states $w_{12}$ [Fig.~\ref{f110}(d)].

The charge redistribution over $z$ direction with the varying gate voltage discussed above should manifest itself in the $V_g$ dependence of the capacitance between the QW and gate electrode.

\begin{figure}
\includegraphics[width=1.0\linewidth,clip=true]{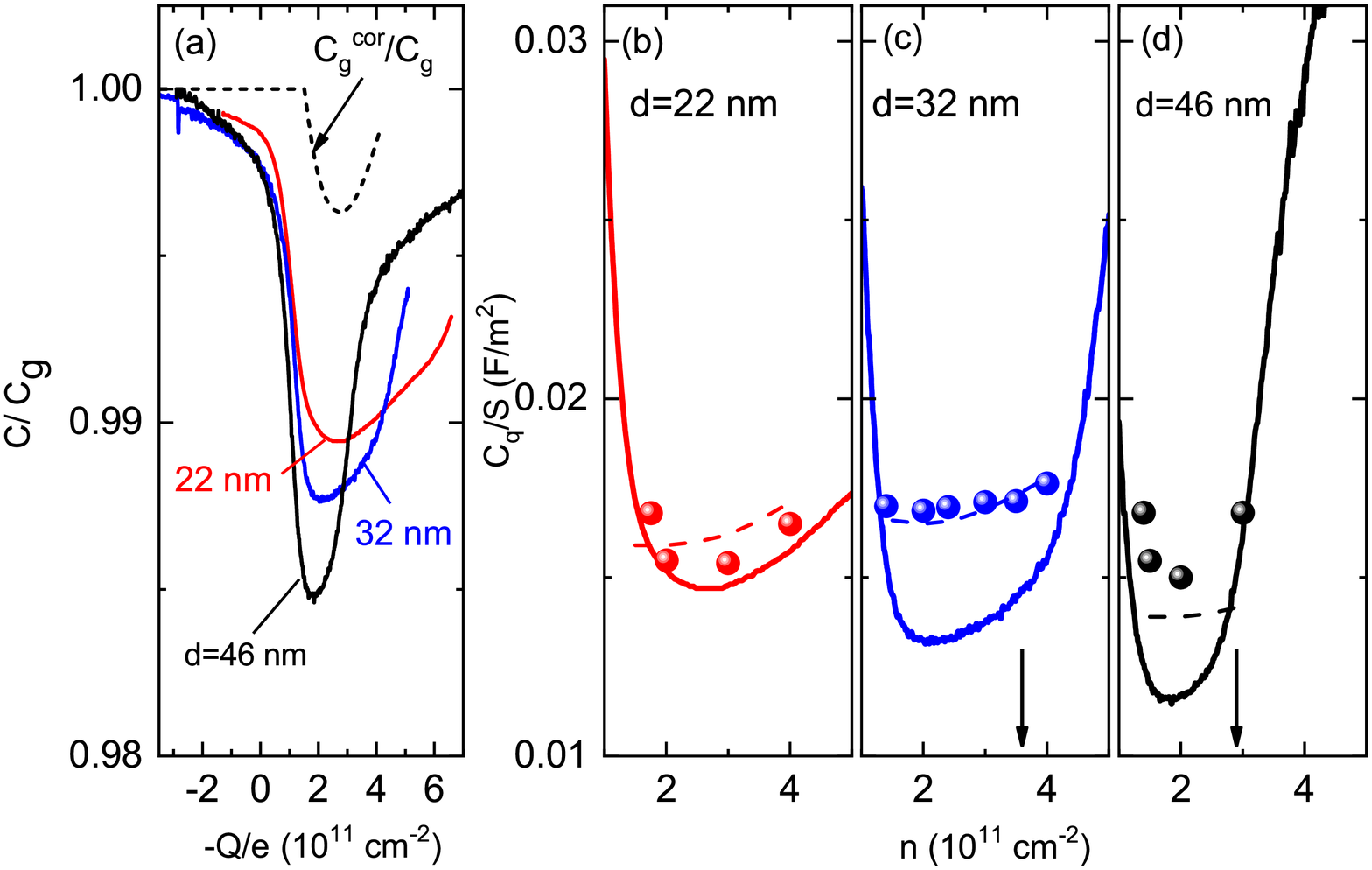}
\caption{(Color online) (a) The  dependence of  $C/C_{g}$ on the charge of the free carriers for different QWs.  The dashed curve is an example for QW with $d=32$~nm of the corrected geometrical capacity which provides the coincidence of the quantum capacity found from the experimental dependence $C(V_g)$ [the dashed line in the panel (c)] with that obtained from the density of states with use of the experimental effective mass [the balls in the panel (c)].  (b) -- (d)  The solid lines are the quantum capacity $C_q / S$ obtained as described in the text with $C_{g}=const$. The balls are the $C_q /S$ values calculated with  $m$ value measured on the same structures (see Fig.~\ref{f6}). The dashed lines  are $C_q /  S$ obtained with taking into account the $V_g$ dependence of the geometrical capacity. The arrows show the electron density at which the second subband of spatial quantization starts to be occupied.}
\label{f12}
\end{figure}

\section{The results of capacitance measurements}
\label{sec:capac}

The gate voltage dependence of capacitance between 2D gas and the gate electrode   results from the finite density of states of a 2D gas and it can be written as:
\begin{equation}
C^{-1}=C_g^{-1}+C_q^{-1},\,\,\,     C_q=e^2D,
\label{eq40}
\end{equation}
where $C_g$ is geometrical capacitance, $D$ is the density of states of 2D gas which is related to the effective mass of the carriers at the Fermi level as $D=m/(\pi\hbar^2 )$.

The samples under study  differ slightly (by $20$~\%) in dielectric thickness, gate area and $V_g$ value corresponding to the charge neutrality point, therefore  the results of the volt-capacitance measurements obtained for different structures are shown in Fig.~\ref{f12}(a) as  $C/C_g$~vs~$Q/e$ plot.

To determine $C_g$, we used the fact that the measurements of the Hall density and SdH oscillations at different $V_g$ and $T$ in the ``hole'' region showed that (i) the degeneracy  of the states of valence band top  is equal to two up to the hole density $(4-5)\times 10^{11}$~cm$^{-2}$; (ii) the effective mass of holes is large, its value is close to each other for all the structures and is equal to $m_h = (0.25 \pm 0.05) m_0$. With taking this into account the $C_g$ value can be obtained  as $C_g = (1 / C - 1 / C_q^{holes})^{-1}$, where $C$ is the experimental capacity measured within the hole region at $p = (2-3)\times 10^{11}$~cm$^{-2}$, $C_q^{holes}=e^2 m_h/(\pi\hbar^2)$ is the hole quantum capacity.

Figure~\ref{f12}(a) shows that  the capacitance drops sharply $n=(0.5-1.0)\times 10^{11}$~cm$^{-2}$ and has a minimum at the nonzero electron densities $n=(1.9-2.5)\times 10^{11}$~cm$^{-2}$. This is due to the two factors: (i) at these QW widths, the conduction and valence bands  overlap; (ii) the effective mass of holes is much (about $10$ times) larger than that of the electrons. In ordinary structures, the drop is associated with a decrease in the density of states and, therefore, with a large contribution of the quantum capacitance. The values of quantum capacity $C_q / S$ calculated from these dependences with a fixed geometric capacity are shown in Figs.~\ref{f12}(b)--\ref{f12}(d). It is seen that the $C_q / S$ value at the minimum decreases with an increase in the QW width, which seems to correspond to a smaller value of the electron mass in wider QWs. However this contradicts the experimental results on the electron masses obtained from the analysis of SdH oscillations shown in Fig.~\ref{f6}.

In order to understand what the seemingly smaller value of the electron mass found from $C(V_g)$ can be connected with, let us consider what approximations were made when obtaining the formula, Eq.~(\ref{eq40}). It was tacitly assumed that the charge distribution in QW over $z$ axis does not change with a change in the density of charge carriers by the gate voltage. Recall that self-consistent calculations (see Fig.~\ref{f10} and Fig.~\ref{f11}) show that $V_g$ changes the charge distribution in $z$ direction. In our case, as discussed above, electrons are repulsed from the gate electrode at a positive voltage that can be considered as the change (decrease in the given case) of the geometrical capacity.

\begin{figure}
\includegraphics[width=0.7\linewidth,clip=true]{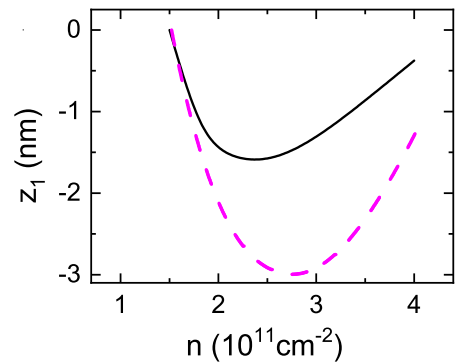}
\caption{(Color online) (a) The solid line is the  electron density dependence of the center of gravity of the electron charge distribution $z_{1}$ for the structure with $d=32$~nm. The dashed line is obtained from $C_g^{cor}$~vs~$Q/e$ dependence shown in Fig.~\ref{f12}(a) with the use of the formula of the flat capacitor.    }
\label{f13}
\end{figure}

Using this simple model one can easily find how the geometrical capacity $C_g^{cor}$ should depend on the gate voltage so that $C_q$ calculated from Eq.~(\ref{eq40}) with replacement of $C_g$ by $C_g^{cor}$ coincides with $C_q$ calculated from the density of states with the use of the experimental effective mass. The results for $C_g^{cor}(V_g)$ for one of the structures is shown in Fig.~\ref{f12}(a) by the dashed line. As  Fig.~\ref{f12}(a) and Fig.~\ref{f12}(c) illustrate the decrease of the geometrical capacity by the value of $0.3$\% in the minimum for structure with $d=32$~nm gives good agreement between $C_q$ values calculated from the experimental effective mass and found from the $C$ vs $V_g$ measurements. The simple estimate with the use of the formula of the flat capacitor gives that this decrease in the geometrical capacitance  corresponds to increase of the distance between the capacitor plates, i.e., to the shift of the electron density away from the gate electrode, by the value of about $3$~nm (Fig.~\ref{f13}). This simple consideration agrees qualitatively with the results of theoretical calculations. This is illustrated by the same figure in which we also show the electron density dependence of the  position  the center of gravity of the electron charge distribution ($z_{1}$) relative to the QW center.

Thus, the redistribution of charge with the gate voltage in wide QWs is significant, and it is this redistribution that changes the $C(V_g)$ dependence and results in the growing MISO amplitude with the electron density increase.

\section{Conclusion}
\label{sec:concl}

We have experimentally studied the magnetotransport and capacitance between 2D gas and the gate electrode in HgTe-based quantum wells of $(20.2-46.0)$~nm width. The data obtained were analyzed using the well-known approach of the self-consistent calculation of the energy spectrum  within the framework of the four-band \emph{kP} model.

It has been shown that the splitting of the conduction band spectrum occurs due to the spin-orbit interaction, therewith, the split branches are single-spin. The splitting strength increases with the increase of the QW width. The electron effective masses in the branches at the Fermi energy  are close to each other  within the actual density range.

In all the structures under study, magneto-intersubband oscillations (MISO) are observed. This indicates that the overlap of the wave functions is sufficiently large. The MISO amplitude in QWs with  $d> 30$~nm increases with an increase of the total electron density. This contradicts to the expected decrease of wave function overlap with increasing $n$ calculated for the rectangular QW.  To interpret such a behaviour, we applied self-consistent approach to calculate the QW energy spectrum. It has been shown that the MISO amplitude increase results from the increasing overlap of the wave functions  due to their shift with the gate voltage  increase which, in its turn, leads to an increase in the probability of transitions between the branches and to an increase of the  MISO amplitude.

Analyzing the data we conclude that standard  self-consistent calculations  are only in qualitative agreement with the experimental results. A possible reason is the difference of dielectric susceptibility of think layer forming the quantum well from that of the parent bulk material as discussed in Refs.~\cite{Andlauer09,Zhu18,Bruene14,Ziegler20}. We are not aware of papers in which this was considered for such a complicated spectrum as HgTe/Hg$_{1-x}$Cd${_x}$Te QWs. Further theoretical studies are needed to understand which effects are not taken into account yet in order to describe the data quantitatively.

\begin{acknowledgments}
The research was supported by a grant of Ministry of Science and Higher Education of the Russian Federation No. 075-15-2020-797 (13.1902.21.0024).
\end{acknowledgments}


%

\end{document}